\documentclass[conference]{IEEEtran}
\IEEEoverridecommandlockouts
\usepackage{cite}
\usepackage{amsmath,amssymb,amsfonts}
\usepackage{graphicx}
\usepackage{tikz}
\usepackage{pgfplots}
\usepackage{pgfplotstable}
\pgfplotsset{compat=1.7}
\usepackage[font=footnotesize]{subcaption}
\usepackage[font=footnotesize]{caption}
\usepackage{textcomp}
\usepackage{url} 
\def\BibTeX{{\rm B\kern-.05em{\sc i\kern-.025em b}\kern-.08em
    T\kern-.1667em\lower.7ex\hbox{E}\kern-.125emX}}
\usepackage{amsmath}
\usepackage{amssymb}
\usepackage{optidef}
\usepackage{algorithm}
\usepackage{xcolor}
\usepackage{algpseudocode}

\newcommand*\Ex[1]{\mathbb{E}\begin{Bmatrix} #1 \end{Bmatrix}}

\newcommand*\mb[1]{\mathbf{#1}}
\newcommand*\bs[1]{\boldsymbol{#1}}

\usepackage{geometry}
\pgfplotsset{every axis/.append style={
		scaled x ticks = false,
		label style={font=\footnotesize},
		tick label style={font=\footnotesize},
		tick scale binop=\times
	}
}
\pgfkeys{/pgf/number format/.cd,
	1000 sep={},
}

\title{Location-Based Load Balancing for Energy-Efficient Cell-Free Networks}
\author{\IEEEauthorblockN {Robbert Beerten\IEEEauthorrefmark{1}, Vida Ranjbar\IEEEauthorrefmark{1}, Andrea P. Guevara\IEEEauthorrefmark{1}, Hazem Sallouha\IEEEauthorrefmark{1}, Sofie Pollin\IEEEauthorrefmark{1}\IEEEauthorrefmark{2}}
\IEEEauthorblockA{\IEEEauthorrefmark{1}\textit{Department of Electrical Engineering, KU Leuven, Belgium}}
\IEEEauthorblockA{\IEEEauthorrefmark{2}\textit{IMEC, Kapeldreef 75, 3001 Leuven, Belgium}}
\IEEEauthorblockA{Corresponding Author: \{robbert.beerten\}@kuleuven.be}
\thanks{This research has received funding from the European Union’s Horizon 2020 research and innovation programme under Grant Agreement No. 101017171 (MARSAL), Horizon Europe under grant agreement No. 101096954 (6G-BRICKS) and from Research Foundation – Flanders (FWO) project number G0C0623N. Hazem Sallouha's work was funded by FWO under the Postdoctoral Fellowship No. 12ZE222N.}}

\begin{document}
\maketitle%
\begin{abstract}
    Cell-Free Massive MIMO (CF mMIMO) has emerged as a potential enabler for future networks.
    It has been shown that these networks are much more energy-efficient than classical cellular systems when they are serving users at peak capacity.
    However, these CF mMIMO networks are designed for peak traffic loads, and when this is not the case, they are significantly over-dimensioned and not at all energy efficient. 
    To this end, Adaptive Access Point (AP) ON/OFF Switching (ASO) strategies have been developed to save energy when the network is not at peak traffic loads by putting unnecessary APs to sleep. Unfortunately, the existing strategies rely on measuring channel state information between every user and every access point, resulting in significant measurement energy consumption overheads. Furthermore, the current state-of-art approach has a computational complexity that scales exponentially with the number of APs. In this work, we present a novel convex feasibility testing method that allows checking per-user Quality-of-Service (QoS) requirements without necessarily considering all possible access point activations. We then propose an iterative algorithm for activating access points until all users' requirements are fulfilled. We show that our method has comparable performance to the optimal solution whilst avoiding solving costly mixed-integer problems and measuring channel state information on only a limited subset of APs.
\end{abstract}
\section{Introduction}
Cell-Free Massive MIMO (CF mMIMO) has recently emerged as
a promising architecture to satisfy the increasing demand for higher spectral efficiency and to address the inadequate performance for cell-edge users
 \cite{Demir_Björnson_Sanguinetti_2021}. In CF mMIMO, densely deployed
Access Points (APs) jointly serve users via coherent transmission and reception \cite{Ngo_Ashikhmin_Yang_Larsson_Marzetta_2017}. The dense deployment of APs, which generally outnumber the users, ensures an improved spectral efficiency performance with uniform service and exceptional
coverage compared to cellular AP deployment \cite{Nayebi_Ashikhmin_Marzetta_Yang_Rao_2017}.
In addition, recent works demonstrated that CF mMIMO is significantly more 
energy-efficient than classic cellular architectures in high-traffic scenarios \cite{Ngo_Tran_Duong_Matthaiou_Larsson_2018} due to smaller AP-user distances and, hence, reduced path losses.

Most existing works design their respective networks for peak capacity. However, it is well-known that traffic load in broadband networks is highly variable with strong periodic behaviour \cite{Piovesan_Lopez-Perez_De_Domenico_Geng_Bao_Debbah_2022, Auer_Giannini_Desset_Godor_Skillermark_Olsson_Imran_Sabella_Gonzalez_Blume_etl}.
Such a variable traffic load renders CF mMIMO networks no longer energy efficient at all when operating below their peak capacity \cite{Viswanathan_Wesemann_Du_Holma, Auer_Giannini_Desset_Godor_Skillermark_Olsson_Imran_Sabella_Gonzalez_Blume_etl}, raising concerns about the overall energy consumption and the corresponding energy-related pollution. 


To address the power-consumption concerns in CF mMIMO networks, AP ON/OFF Switching (ASO) strategies are attracting considerable research focus \cite{Van_Chien_Björnson_Larsson_2020,Sallouha_Sarkar_Krijestorac_Cabric_2023}. The core idea of ASO strategies is to put the underused set of APs entirely in a deep sleep state. Given that the power consumption of APs consists mainly of the power consumed at the Power Amplifier (PA) and the digital processing chain \cite{Viswanathan_Wesemann_Du_Holma}, a fundamental trade-off comes with ASO strategies. On the one hand, by putting APs to sleep, one can eliminate their fixed power consumption; however, on the other hand, putting APs to sleep leads to increased power consumption at the PA of active APs as fewer APs typically result in larger AP-user distances and, hence, path losses.
Herein, the goal is to find the optimal point between the fixed power consumption of APs and the 
power consumed at said APs' PAs.
Achieving this goal requires solving an optimization problem, which can be formulated as the 
joint consumption of fixed power at each AP and the total power consumption of all PAs under 
per-user Signal-to-Interference-and-Noise-Ratio (SINR) requirements and a total transmit power constraint per AP \cite{Nayebi_Ashikhmin_Marzetta_Yang_Rao_2017}. 
The objective of such an optimization problem can be formulated as a mixed-binary second-order cone program (MB-SOCP) and 
is proven to be convex in \cite{Van_Chien_Björnson_Larsson_2020}. However, this problem is computationally intensive due to the 
involvement of the aforementioned binary variables related to the potential activation of APs. 

Binary variables, representing the optimal set of APs to be activated, are usually found by traversing a binary decision tree via branch-and-bounding approaches.
The size of the branch-and-bound tree then scales exponentially with the considered number of APs, which 
is generally an order of magnitude larger than the number of users in a CF mMIMO network. 
This discrepancy is even more significant if the network operates significantly below its' maximum 
traffic load, i.e., users request less traffic or a smaller percentage of users is active.
This complexity renders the optimization procedure 
computationally infeasible for any realistic network size and surely for the targeted use case in which the APs significantly outnumber the users.
To reduce the computational complexity of the ASO optimization problem, current state-of-the-art solutions proposed minimizing only the number of active APs as opposed to minimizing the total power consumption \cite{Van_Chien_Björnson_Larsson_2020, Demir_Masoudi_Björnson_Cavdar_2023}.
The existing state-of-the-art strategies rely on measuring channel state information between every user and every access point, resulting in significant measurement energy consumption overheads. To the best of our knowledge, the need for a computational-efficient and overhead-efficient ASO optimization solution is an open research question. 
\noindent
\\
\textbf{Contribution:}
This work goes beyond the current state-of-the-art by only activating APs for channel measurements if it is proven that it would be otherwise infeasible to satisfy users' SINR requirements.
This allows to leave those sleeping APs in a deep sleep state, resulting in significantly more energy saving \cite{3gpp.38.864}.
A framework to prove the infeasibility of such a problem via slack variables is provided.
It is then proven that minimizing these slack variables is a convex problem. 
These slack variables are then used to design a per-user location-based heuristic scheme to rapidly find a subset of APs that can satisfy each user's Downlink (DL) SINR requirements. 
This leads to a cheap solution close to the optimal solution of the MB-SOCP.
We summarize our contribution in three points:
\begin{itemize}
    \item We construct a slack variable-based framework that enables a detailed analysis of per-user SINR constraints, which only depends on the active APs. 
    \item We propose a low-complexity AP switching strategy that avoids solving expensive mixed-integer optimization problems based on our proposed slack variable-based framework. We then propose to activate extra APs for channel measurement only if needed instead of computing the solution for each possible subset of APs. 
    \item We show that our method performs close to the optimal solution in \cite{Van_Chien_Björnson_Larsson_2020} while being considerably lower in complexity and requiring significantly fewer APs to be active for measuring the channel state information when solving the problem compared to the optimal solution.
    \end{itemize} 
%
\textbf{Notation:}
 A diagonal matrix with the elements $x_i$ on the main diagonal is denoted by $\text{diag}(x_0, x_1, \cdots, x_N)$.
 The cardinality of a set $\mathcal{S}$ is written as $|\mathcal{S}|$. For matrices and vectors alike, $\mathbf{A}^T$ is the transpose,
 $\mathbf{A}^H$ is the Hermitian conjugate of the matrix, and $\mathbf{A}^{\ast}$ is the complex conjugate. 
 The $l$'th element of vector $\mathbf{a}$ is denoted by $[\mathbf{a}]_l$. The expected value of $x$ is denoted by 
 $\Ex{x}$. A real multivariate normal distribution, with covariance $\mb{A}$ is denoted by $\mathcal{N}(\mb{0}, \mb{A})$, its complex counterpart by $\mathcal{CN}(\mb{0}, \mb{A})$.
 The notations $\mathbb{R}^{M \times N}, \mathbb{R}^{N}_{\succeq \mb{0}}$ and $\mathbb{C}^{M \times N}$ denote the set of real $M$ by $N$ matrices, real $N$ dimensional vectors with non-negative elements and complex $M$ by $N$ matrices, respectively.

\section{System Model}
\label{sec:sysmodel}
We consider a system with $K$ single-antenna users, $L$ APs with $N$ antennas each, and a single 
CPU. 
The wireless channel between every user $k$ and AP $l$ is considered to be Rayleigh fading and is thus modelled as a realization of a zero-mean multivariate random distribution as 
$\mathbf{h}_{lk} \sim \mathcal{CN}(\mb{0}, \mb{R}_{lk})$.
The channel covariance matrix $\mb{R}_{lk} \in \mathbb{C}^{N\times N}$ is generated according to \cite[Sec. 2.6]{Björnson_Hoydis_Sanguinetti_2017}. 
The channel is assumed to be reciprocal in the uplink (UL) and DL.
Each user has a distinct pilot sequence. 
The UL channel is estimated at every AP as
\begin{equation}
        \hat{\mb{h}}_{lk} = \mb{R}_{lk} (\mb{R}_{lk} + \sigma_{UL}^2\mb{I}_N)^{-1} \mathbf{y}^{(p)}_{lk}\,,
\end{equation}
where $\sigma_{UL}^2$ is the thermal noise at AP $l$ for user $k$ and $\mathbf{y}^{(p)}_{lk}$ is the decorrelated pilot symbol at AP $l$ for user $k$ \cite{Björnson_Hoydis_Sanguinetti_2017}. To reduce the computational complexity in each AP, each AP uses maximum-ratio beamformers in the DL $ \mb{w}_{l,k}$, which are defined as
\begin{equation}
    \mb{w}_{l,k} = \frac{\hat{\mb{h}}_{lk}}{\|\hat{\mb{h}}_{lk}\|_2}.
\end{equation}
The received signal at user $k$ can then be written as, 
\begin{equation}
\begin{aligned}
    y_k =& \sum_{l=1}^L  \sum_{i=1}^{K} \sqrt{\rho_{li}} \mb{h}_{lk}^H \mb{w}_{li} s_i + n_k \\
    y_k =& \sum_{l=1}^L\sqrt{\rho_{lk}}  \Ex{\mb{h}_{lk}^H \mb{w}_{lk}}  s_k  + \sum_{l=1}^L\sum^{K}_{i=1} \sqrt{\rho_{li}} \mb{h}_{lk}^H \mb{w}_{li}  s_i \\
    &  - \sum_{l=1}^L\sqrt{\rho_{lk}}  \Ex{\mb{h}_{lk}^H \mb{w}_{lk}} s_k  + n_k,
    \end{aligned}
    \end{equation}
where $\rho_{lk}$ is the DL power allocated to user $k$ at AP $l$, $n_k$ is the thermal
noise at user $k$ and is generated from the distribution $\mathcal{CN}(0, \sigma_{DL}^2)$, the transmitted signal $s_k$ is assumed to be unit power and is mutually uncorrelated amongst different users. 
The use-and-then-forget bound on the DL SINR for user $k$ is then \cite{Demir_Björnson_Sanguinetti_2021}
\begin{equation}
\begin{aligned}
    & \gamma_k =\\ 
    &\frac{| \sum\limits_{l=1}^L\sqrt{\rho_{lk}} \Ex{\mb{h}_{lk}^H \mb{w}_{lk}} |^2}
    {\sum\limits_{i=1}^K \Ex{|\sum\limits_{l=1}^L \sqrt{\rho_{li}} \mb{h}_{lk}^H \mb{w}_{li}|^2} - |\sum\limits_{l=1}^L \sqrt{\rho_{lk} }\Ex{\mb{h}_{lk}^H \mb{w}_{lk}}|^2 + \sigma_{DL}^2}.
    \end{aligned}
\end{equation}
Which can be rewritten as,
\begin{equation}
    \gamma_k = \frac{|\mb{b}_k^T\bs{\rho}_{k}|^2}{\sum\limits_{i=1}^K\bs{\rho}_i^{T}\mathbf{C}_{ki}\bs{\rho}_i - |\mb{b}_k ^{T}\bs{\rho}_{k}|^2 + \sigma_{DL}^2} \,,
    \label{eqn:ErgRate}
\end{equation}
where the vector $\bs{\rho}_k = [\sqrt{\rho_{1k}} \; \sqrt{\rho_{2k}} \; \dots \; \sqrt{\rho_{Lk}}]^T$ contains the DL power coefficients between the $k$-th user and each AP. Moreover, the symbols $\mb{b}_k  \in \mathbb{R}_{\succeq \mb{0}}^{L} $
and $\mb{C}_{ki}  \in \mathbb{C}^{L \times L}$, which are defined as,
\begin{subequations}
    \begin{align}
        [\mathbf{b}_k]_l &=  \text{Re}\{ \Ex{ \mb{h}_{lk}^H \mb{w}_{lk}} \}\,, \\ 
        [ \mathbf{C}_{ki}]_{l,m} &= \Ex{\mb{h}_{lk}^{H}\mb{w}_{li}\mb{w}_{mi}^{H} \mb{h}_{mk}},
    \end{align}
\end{subequations}
are introduced for notational brevity. 
To study the case where only a limited subset of APs are active, which is denoted by $\mathcal{A}$, the following two notations are also introduced: $\bar{\mathbf{b}}_k \in \mathbb{R}_{\succeq \mb{0}}^{|\mathcal{A}|}$ and $ \bar{\mathbf{C}}_{ki} \in \mathbb{C}^{|\mathcal{A}| \times |\mathcal{A}|}$,
\begin{subequations}
    \begin{align}
    [\bar{\mathbf{b}}_k]_l &= \text{Re} \{ \Ex{\mb{h}_{lk}^H \mb{w}_{lk}}\}  \; \;\qquad & \forall l \in \mathcal{A}\,, \\ 
    [ \bar{\mathbf{C}}_{ki}]_{l,m} &= \Ex{\mb{h}_{lk}^{H}\mb{w}_{li}\mb{w}_{mi}^{H} \mb{h}_{mk}}  &\forall l,m \in \mathcal{A}\,,
    \end{align}
\end{subequations}
which only consider the non-zero elements of both $\mb{b}_k$ and $\mb{C}_{ki}$. This shall be used in Section~\ref{subsec:alloc} to reduce 
the dimensionality of the SINR constraints. Consequently, $\bar{\bs{\rho}}_k$ is then the vector of DL power coefficients for APs in $\mathcal{A}$, which can be directly used to estimate the DL SINR for user $k$ as,
\begin{equation}
       \gamma_k = \frac{|\bar{\mb{b}}_k^T\bar{\bs{\rho}}_{k}|^2}{\sum\limits_{i=1}^K\bar{\bs{\rho}}_i^{T}\bar{\mathbf{C}}_{ki}\bar{\bs{\rho}}_i - |\bar{\mb{b}}_k ^{T}\bar{\bs{\rho}}_{k}|^2 + \sigma_{DL}^2}.
\end{equation}
For the rest of this paper, the term Spectral Efficiency (SE) refers to the DL ergodic rate, which can be expressed as $\mathrm{SE}_k = \log_2(1+\gamma_k)$.
This relation is important as we will use SINR and SE interchangeably throughout this paper, depending on the context.
The optimization problems defined in Sections \ref{subsec:alloc} and \ref{sec:feas} have an SINR constraint. For the results in Section~\ref{sec:res}, using the 
SE to dimension the constraint is more tractable.

\subsection{Power Consumption Model}
This work adopts the power consumption model presented in \cite{Van_Chien_Björnson_Larsson_2020}. The total consumed power for a given set of active APs, $\mathcal{A}$, is given by,
\begin{equation}
     P_{\mathcal{A}} = P_{BB} \sum_{l \in \mathcal{A}} x_l +  \Delta  \sum_{l \in \mathcal{A}}  \sum_{k=1}^{K} \rho_{lk} \,,
\end{equation}
where $P_{BB}$ and $x_l$ are the fixed power consumption of an active AP and the binary variable indicating if an AP is active, respectively. The PAs' power consumption inefficiency is expressed via $\Delta$. Each PA is also subject to a maximum power transmission constraint, denoted by $P_{max}$.

\subsection{Downlink Power Allocation}
\label{subsec:alloc}
We introduce the following optimization problem, $\mathcal{P}1$, via the convex formulation proposed in \cite{Van_Chien_Björnson_Larsson_2020} for uncorrelated Rayleigh fading and subsequently extended in \cite{Demir_Björnson_Sanguinetti_2021} for correlated Rayleigh fading,
\begin{mini!}[3]
 	{\rho_{l,k} \geq 0, x_{l} \in \{0, 1\}}
    {\hspace{-0.5cm}P_{BB} \sum_l x_l + \Delta \sum_{l=1}^{L}\sum_{k=1}^{K}   \rho_{l,k}} 
    {\label{p:MISOCP}}
    {\mathcal{P}1:\hspace{-0.8cm}}  
    \addConstraint
    {\bar{\gamma}_k}
    {\leq \frac{|\mb{b}_k^T\bs{\rho}_{k}|^2}{\sum\limits_{i=1}^K\bs{\rho}_i^{T}\mathbf{C}_{ki}\bs{\rho}_i - |\mb{b}_k ^{T}\bs{\rho}_{k}|^2 + \sigma_{DL}^2}\qquad \label{eq:SEconstraint}}
    {\forall k}
    \addConstraint
    {\sqrt{P_{max}} x_l }
    {\geq \| \sqrt{\rho_{l1}} \; \dots \; \sqrt{\rho_{lK}} \|_2 \label{con:mixer}}
    {\forall l}\,.
\end{mini!}
This problem can then be cast as an MB-SOCP via the epigraph trick on the cost function \cite[Lemma 3]{Van_Chien_Björnson_Larsson_2020}. In this problem, the binary variables $x_l$ indicate if AP $l$ is active. Constraint~(\ref{eq:SEconstraint}) ensures that each user's SINR is at least $\bar{\gamma}_k$ and constraint (\ref{con:mixer})
ensures consistency between the binary variables and the power coefficients while ensuring that per PA power consumption does not exceed the limit.
The binary variables in this problem are problematic as an exhaustive search over all combinations of these variables would require solving $2^{L}$ SOCPs, and thus, the problem scales exponentially.
This problem can be solved optimally and more efficiently
via Branch-and-Bound (B\&B). Unfortunately, a thorough analysis of computational cost for B\&B is an open problem, and we have to rely on the exhaustive upper bound ($2^{L}$ SOCPs to be solved), which tends to be rather loose. 
The B\&B algorithm is considered to be the baseline for our proposed algorithm.
This algorithm is executed at the network's central processing unit (CPU).
Note that the constraint (\ref{eq:SEconstraint}) can be formulated as a second-order cone constraint via the following formulation \cite{Van_Chien_Björnson_Larsson_2020}:
\begin{align}
\sqrt{\frac{1 + \bar{\gamma}_k }{\bar{\gamma}_k }}\mb{b}_k ^{T}\bs{\rho}_{k}
    & \geq
    \| [ (\mathbf{C}_{k1}^{\frac{1}{2}}\bs{\rho}_1)^T \: \dots \:  (\mathbf{C}_{kK}^{\frac{1}{2}}\bs{\rho}_K)^{T}  \; \sigma_{DL}]^T \|_2    \nonumber\\
    & \geq   \mathcal{I}_k(\bs{\rho}_1, \dots, \bs{\rho}_K) \,,
     \label{eqn:inter1}
    \end{align}
where we introduce $\mathcal{I}_k(\bs{\rho}_1, \dots, \bs{\rho}_K)$ as the interference term towards user $k$ for notational simplicity. For a limited set $\mathcal{A}$, we use the following notation:
\begin{equation} 
 \bar{\mathcal{I}_k}(\bar{\bs{\rho}}_1, \dots, \bar{\bs{\rho}}_K) = \| [ (\bar{\mathbf{C}}_{k1}^{\frac{1}{2}}\bar{\bs{\rho}}_1)^T \: \dots \:  (\bar{\mathbf{C}}_{kK}^{\frac{1}{2}}\bar{\bs{\rho}}_K)^{T}  \; \sigma_{DL}]^T \|_2.
 \label{eqn:inter2}
 \end{equation}
 Note that the previous notation does not specify $\mathcal{A}$. In this section, $\mathcal{A}$ is assumed to be known; section~\ref{sec:algo} will expose the exact construction of this set.
 Following  (\ref{eqn:inter1}) and (\ref{eqn:inter2}), the following power  minimization problem $\mathcal{P}2$ is defined for a given $\mathcal{A}$, 
\begin{mini!}[3]
 	{\rho_{l,k} \geq 0}
    {\sum_{l \in \mathcal{A}} \sum_{k=1}^K   \rho_{l,k}} 
    {}
    {\mathcal{P}2:}  
    \addConstraint
    {\sqrt{\frac{1 + \bar{\gamma}_k }{\bar{\gamma}_k }}\bar{\mb{b}}_k ^{T}\bar{\bs{\rho}}_{k}}
    { \geq \bar{\mathcal{I}_k}(\bar{\bs{\rho}}_1, \dots, \bar{\bs{\rho}}_K) \qquad \label{con:noSlack}}
    {\forall k}
    \addConstraint
    {\sqrt{P_{max}} }
    {\geq \| \sqrt{\rho_{l1}} \; \dots \; \sqrt{\rho_{lK}} \|_2 \label{con:aff2}}
    {\forall l \in \mathcal{A}}
\end{mini!}
This problem reduces the dimensions of the conic constraints due to the reduced dimensions of $\bar{\mathbf{C}}_{ki}$'s, $\bar{\mb{b}}_k$'s and $\bar{\bs{\rho}}_k$'s and does not contain binary variables. Note that the result of this procedure requires the addition of the fixed power consumption per AP, which equates to  $|\mathcal{A}|P_{BB}$.

\section{Feasibility Problem}
\label{sec:feas}
Problem $\mathcal{P}3$ is introduced to check if the currently active APs are sufficient to satisfy the users' SINR requirements.  
This is a so-called feasibility testing problem \cite[Sec. 11.4]{Boyd_Vandenberghe_2004}. Due to the addition of slack variables, $s_k$,
on the per-user SINR constraints (\ref{con:slack}), this problem is always feasible. More importantly, the values of the slack variables at the solution of this problem provide insight into which QoS requirements in the original problem $\mathcal{P}2$ cannot be satisfied.
\begin{mini!}[3]
 	{\rho_{l,k} \geq 0, s_k}
    {\sum_{k=1}^{K} s_k} 
    {}
    {\mathcal{P}3:}  
    \addConstraint
    {s_k}
    {  \geq \bar{\mathcal{I}_k}(\bar{\bs{\rho}}_1, \dots, \bar{\bs{\rho}}_K) - \sqrt{\frac{1 + \bar{\gamma}_k }{\bar{\gamma}_k }}\bar{\mb{b}}_k^T\bs{\bar{\rho}}_{k} \;\label{con:slack}}
    {\forall k}
    \addConstraint
    {s_k}
    {\geq 0 \label{con:aff1} \qquad}
    {\forall k}
    \addConstraint
    {\sqrt{P_{max}} }
    {\geq \| \sqrt{\rho_{l1}} \; \dots \; \sqrt{\rho_{lK}} \|_2 \label{con:aff2}}
    {\forall l \in \mathcal{A}}
\end{mini!}
The slack variables $s_k$ take on the value $\text{max}(0, \bar{\mathcal{I}_k}(\bar{\bs{\rho}}_1, \dots, \bar{\bs{\rho}}_K) - \sqrt{\frac{1 + \bar{\gamma}_k }{\bar{\gamma}_k }}\bar{\mb{b}}_k^T\bs{\bar{\rho}}_{k})$ i.e. they are zero if 
their respective constraint in $\mathcal{P}2$ is satisfied at the solution of $\mathcal{P}3$; otherwise, it is precisely equal to the value of the violation of the constraint in $\mathcal{P}2$ using the solution from $\mathcal{P}3$. This highlights the unique application of the 
slack variable: not only do they provide insight into which constraint cannot be satisfied, but also into which constraints are relatively further from feasibility. Due to the structure of the problem, each of these $s_k$'s corresponds exactly to a user-specific SINR constraint. Thus, given the users' location, we can implement our lightweight heuristic algorithm in Section~\ref{sec:algo}.

The proof that problem $\mathcal{P}3$ is convex is omitted here due to space constraints but can be achieved by reformulating constraint (\ref{con:slack}) as the cone in \cite[Ex 2.11]{Boyd_Vandenberghe_2004}. The aforementioned convexity implies that $\mathcal{P}3$ exhibits only a singular local minimum; thus, the global minimum is always found. Unlike most feasibility testing methods, which start from a fixed
starting point for the variables of the original problem  \cite[Sec. 11.4]{Boyd_Vandenberghe_2004} ($\mathcal{P}2$ in this case), this method searches jointly over both the slack variables and the power coefficients. 

\section{Proposed Algorithm}
\label{sec:algo}
In this section, we present our proposed algorithm for activating APs.
Our algorithm is divided into three distinct steps:
1) \textit{Initial Access}, 2) \textit{Feasible Set Creation} and 3) \textit{AP Pruning}.
In this work, the users' locations are assumed to be known and this information is used to enlarge the set of active APs in case more APs are needed to meet the QoS requirements.
The main advantages of our algorithm are twofold. First, it avoids the traversal of the branch-and-bound tree required for finding the optimal solution. Second, it reduces the dimensions of the intermediate SOCP problems. 

\subsection{Initial Access}
In the initial phase, every user connects to its closest AP, which is now activated from its sleep state. To this end, the users' location is assumed to be known perfectly. This assumption is valid for many use cases, as many devices today are equipped with wireless transceivers for communication and localization. Furthermore, much research is dedicated towards fine-grained user localization based on radio signals \cite{Kanhere_Rappaport_2021}. 
These APs are then the initial set of active APs $\mathcal{A}$ and used for channel measurements, resulting in their respective elements of $\bar{\mb{b}}_{lk}$ and $\bar{\mb{C}}_{ki}$. 

\subsection{Feasible Set Search}
With this initial set of APs $\mathcal{A}$, problem $\mathcal{P}2$ might still be infeasible if the QoS requirement is set sufficiently high. To this end, 
the resulting slack variables of $\mathcal{P}3$ are used. 
Because problem~$\mathcal{P}3$ is always feasible, the global minimum is always found for the slack variables.
If all slack variables are zero at the global minimum, the problem is feasible, and the algorithm terminates.
Otherwise, the set of active APs is enlarged to satisfy the SINR requirements. 
This is achieved by turning on the sleeping AP closest to the user with the largest slack variable in the solution of $\mathcal{P}3$.
If all APs are active and $\mathcal{P}3$ still returns non-zero slack variables, the requested SINR is infeasible for the considered network. 
The pseudocode of the aforementioned steps is provided in Algorithm~\ref{alg:feasible}.
For our simulation, we avoid completely infeasible SINR requirements by first solving the max-min SE problem when every AP in the network is active. 
This procedure returns the highest SINR requirement to be feasible via Problem~$\mathcal{P}1$. 
Measuring channels at only specific APs was also proposed in \cite{Beerten_Ranjbar_Guevara_Pollin_2023} and is extended here to aid in solving the power minimization problem.
\begin{algorithm}[htb]
    \caption{Iterative Feasible Set Finder}
    \label{alg:feasible}
\begin{algorithmic}[1]
    \State{Solve $\mathcal{P}3$ for the initial $\mathcal{A}$}
    \While{$\exists s_k > 0$}
     \If{$|\mathcal{A}| == L$}
            \State{SINR requirement is impossible to satisfy}
            \EndIf
                \State{$k^{\ast} \leftarrow \arg \max_k s_k$}
        \State{Take $l^{\ast}$ as as the closest AP to user $k^{\ast}$}
        \State{$\mathcal{A} \leftarrow ~ \mathcal{A} ~ \bigcup ~ l^{\ast}$ } 
        \State{Solve $\mathcal{P}3$ for the new $\mathcal{A}$}
    \EndWhile
\end{algorithmic}
\end{algorithm}
\subsection{Active AP Pruning}
Algorithm~\ref{alg:feasible} might result in an over-dimensioned network in two cases: 1) the SINR requirement is loose, and the initial access already activates too many APs, and 2) APs activated in later iterations of
Algorithm~\ref{alg:feasible} might render APs from earlier iterations obsolete. These obsolete APs should be put to sleep if they do not reduce the total power consumption.
To this end, we utilize a method similar to the one proposed in \cite{Demir_Masoudi_Björnson_Cavdar_2023}. 
The algorithm checks which AP transmits at the lowest power. 
It then solves the feasibility problem without that AP.
The AP is put to sleep if the problem remains feasible and if it lowers the energy consumption of the network. 
The effect on the energy consumption depends on the relative values of the fixed AP powers, $P_{BB}$, and the power inefficiency coefficient $\Delta$. Pseudocode is provided in Algorithm~\ref{alg:prune}. 

\begin{algorithm}[htb]
    \caption{Prune Active APs}
    \begin{algorithmic}[1]
    \State{Start from intial set $\mathcal{A}$ provided by Algorithm~\ref{alg:feasible}}
    \While{$\mathcal{P}$3 returns a feasible solution}
        \State{$l^{\ast} \leftarrow \arg \min_l \sum_k\rho_{lk}$}
    
        \If{$\mathcal{P}3$ feasible for $\mathcal{A} \setminus l^{\ast} $}
            \State{Solve $\mathcal{P}2$ for $\mathcal{A} \setminus l^{\ast} $}
            \If{$P_{\mathcal{A} \setminus l^{\ast}} < P_{\mathcal{A} }$}
            \State{$\mathcal{A} \leftarrow \mathcal{A} \setminus l^{\ast}  $}
            \EndIf
        \Else
            \State{Terminate Algorithm}
        \EndIf
    \EndWhile
    \end{algorithmic}
    \label{alg:prune}
\end{algorithm}
\subsection{Complexity and Convergence}
The MB-SOCP constructs a binary tree with all possible binary variable combinations in its nodes. Thus, the baseline must solve Problem~$\mathcal{P}2$ $2^L$ times in the worst case. Even though many state-of-the-art branch-and-bounding algorithms can reduce the search space, these provide no guarantee of shorter maximum runtimes \cite{Van_Chien_Björnson_Larsson_2020}, and exponential complexity is still expected. 
Our proposed algorithm must, in the worst case, solve the same Problem~$\mathcal{P}2$ $2L$ times as there are only $L$
APs that can be activated in Algorithm~\ref{alg:feasible} or put to sleep in Algorithm~\ref{alg:prune}. 
Hence making our algorithm significantly more scalable.
The algorithm always finds a solution for any SINR below the max-min rate when the full network is activated, i.e. $|\mathcal{A}| = L$. This is the main strength of the proposed method: in contrast to the baseline, of which the number of SOCPs scales exponentially with the number of APs, is approximated by our method in which the number of SOCPs to be solved scales linearly with the number of APs. 

\section{Results}
\label{sec:res}
This section provides numerical results for our proposed method. 
The users and APs are uniformly distributed in a $500\times500$~m$^2$ area. 
Due to the significant complexity of the baseline \cite{Van_Chien_Björnson_Larsson_2020}, the simulation scenario is kept relatively small.
The parameters of the simulations are listed in Table~\ref{tab:simPars}.
The path loss is modelled as follows,
\begin{equation}
    \text{PL}_{dB} = -35.4 -24\log_{10}(d_{lk}) + F_{shadow},
\end{equation}
where $d_{lk}$ is the distance between user $k$ and AP $l$ (in meters) and $F_{shadow} \sim \mathcal{N}(0,\sigma_{sf}^2)$.
This section provides numerical results both for the number of APs that are active and 
the total power consumption in the network. 
The numerical results are computed using CVX \cite{cvx, gb08} with 
Gurobi \cite{gurobi} in the back-end for solutions to the mixed-integer problem.
The iterative solution is benchmarked against
the solution of the MB-SOCP problem, which is known to be optimal \cite{Van_Chien_Björnson_Larsson_2020}.
\begin{table}[b]
    \centering
    \begin{tabular}{l|r||l|r} 
        Parameter & Value & Parameter & Value\\ \hline \hline
        K & 7 & L & 15 \\ 
        N & 4 & $P_{BB}$ & 5\\
        $\Delta$ & 2 & $P_{max}$ & 1 \\
        $\sigma_{DL}^2$ & -94 dBm & $\sigma_{UL}^2$ & -94 dBm \\ 
        $\sigma_{sf}$ & 4 & B & 20 MHz\\ 
    \end{tabular}
    \caption{Simulation Parameters}
      \label{tab:simPars}
\end{table}
We provide results for three key performance indicators: the number of APs active for measuring their respective channels, the total power consumption of the network, and the energy efficiency. We perform these simulations for a range of SEs between 0.25 and 2.25 Bps/Hz/W as we are mostly interested in the case of a network being over-dimensioned to 
serve its users.

\subsection{Active APs}
\begin{figure}[tb]
\begin{tikzpicture}
\pgfplotstableread[col sep=comma,]{figures/APsRes.csv}\datatable
\begin{axis}[
    xlabel={Target Spectral Efficiency [Bps/Hz]},
    ylabel={Number of Active APs},
    xtick =data,
    legend style={at={(0.05,0.7)},anchor=west},
    legend style={font=\fontsize{7}{5}\selectfont},
    grid style={line width=.1pt, draw=gray!10},
    major grid style={line width=.2pt,draw=gray!50},
    grid style=dashed,
    grid=both,
      width=\linewidth,
    height=0.7\linewidth
]
    \addplot [mark=o, red] table [x, y={y1}]{\datatable};
    \addlegendentry{Proposed Method Measure}
    
    \addplot [mark=x, blue] table [x, y={y2}]{\datatable};
    \addlegendentry{Proposed Method Serve}
    \addplot [mark=diamond, black] table [x, y={y4}]{\datatable};
    \addlegendentry{MB-SOCP \cite{Van_Chien_Björnson_Larsson_2020} Measure}
    
    \addplot [mark=star, brown] table [x , y={y3}]{\datatable};
    \addlegendentry{MB-SOCP \cite{Van_Chien_Björnson_Larsson_2020} Serve}
        
\end{axis}
\end{tikzpicture}
    \caption{Comparison of the number of APs used for measuring and serving for both the optimal 
    solution via the MB-SOCP and our proposed method as a function of the SE requirement.}
    \label{fig:nbAPs}
    \vspace{-0.6cm}
\end{figure}
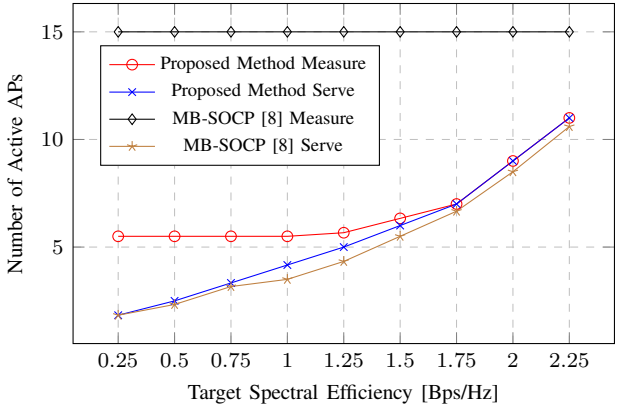
To accurately investigate the number of active APs, we distinguish two notions of active APs. First, the \textbf{Measuring APs} are only used to measure channel statistics and not necessarily to serve users. 
The measuring APs are only activated briefly and then, depending on the output of Algorithm~\ref{alg:prune}, put to sleep if they 
are deemed unnecessary for reaching SE requirements.
Second, the \textbf{Serving APs} are used in the DL transmission.
The optimal solution, reached via the MB-SOCP \cite{Van_Chien_Björnson_Larsson_2020}, requires those measurements 
from every AP in the network. However, a key advantage of our proposed method is that only a small subset of the network needs to be activated for channel measurement. For a target SE of 0.5 Mbps/Hz, we reduce the measuring APs by 64\%, whereas we only increase the number of serving APs by 4\% on average.

In Figure~\ref{fig:nbAPs}, we show the number of measuring and serving APs as a function of the SINR requirement. As illustrated in Figure~\ref{fig:nbAPs}, our solution permits
a considerable subset of APs to be left in a sleep state.  Our proposed method only requires 0.22 APs more on average for the entire range of tested SEs.
 However, our method truly shines when we compare the number of measuring APs. Especially for the range 0.25-1.25 Bps/Hz, our method only uses 5.3 APs on average for measuring, whereas the baseline methods always require the entire network (15 APs) to be active. 
Our method even has the potential to use even fewer APs for measuring by refining the proximity-based initial access procedure.

\subsection{Power Consumption}
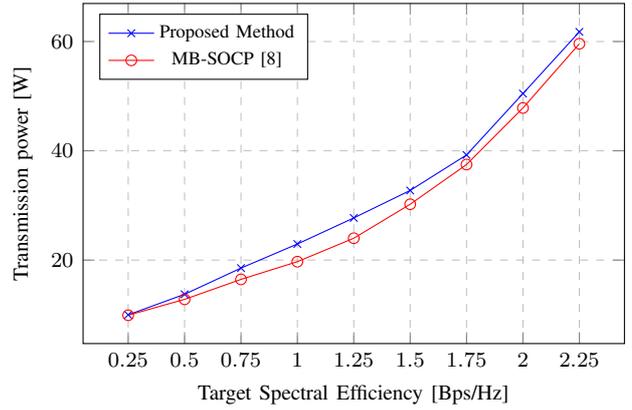
\begin{figure}[t]
\begin{tikzpicture}
\pgfplotstableread[col sep=comma,]{figures/powerRes.csv}\datatable
\begin{axis}[
    xlabel={Target Spectral Efficiency [Bps/Hz]},
    ylabel={Transmission power [W]},
        xtick =data,
    legend pos=north west,
    legend style={font=\fontsize{7}{5}\selectfont},
    grid style={line width=.1pt, draw=gray!10},
    major grid style={line width=.2pt,draw=gray!50},
    grid style=dashed,
    grid=both,
      width=\linewidth,
    height=0.7\linewidth
]
    \addplot [mark=x, blue] table [x, y={y2}]{\datatable};
    \addlegendentry{Proposed Method}
    \addplot [mark=o, red] table [x, y={y1}]{\datatable};
        \addlegendentry{MB-SOCP \cite{Van_Chien_Björnson_Larsson_2020}}

\end{axis}
\end{tikzpicture}
    \caption{The effect of the SINR requirement on the total power consumption for DL transmission when serving users.}
    \label{fig:power}
            \vspace{-0.4cm}
\end{figure}

Figure~\ref{fig:nbAPs} has shown that both methods use a similar number of APs for serving users, but we do not know if both selected subsets also consume comparable power. We thus investigate if our selected subset of APs reaches a power consumption close to the optimal as computed by \cite{Van_Chien_Björnson_Larsson_2020}. Since our method only activates a small subset of APs for measuring the channel, the goal is to get as close to the minimum power consumption as computed by the baseline. As evidenced by Figure~\ref{fig:power}, our method consumes only marginally more power, 7\% on average, than the MB-SOCP solution for a range of scenarios with 15 APs and seven users for the entire range of possible SEs and only 12\% in the worst case (1 Bps/Hz). 
Thus, we are very close to the optimal solution while being significantly cheaper in computational cost and measurement power.
This underlines our intuition that simple heuristics can provide comparable performance to the baseline.

\subsection{Energy Efficiency}
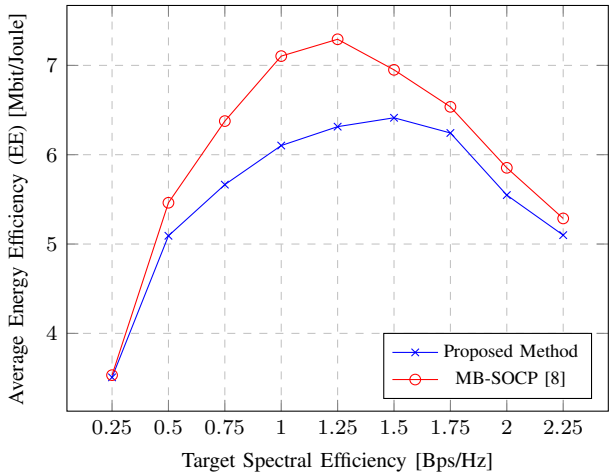
\begin{figure}[htb]
\begin{tikzpicture}
\pgfplotstableread[col sep=comma,]{figures/energEff.csv}\datatable
\begin{axis}[
    xlabel={Target Spectral Efficiency [Bps/Hz]},
    ylabel={Average Energy Efficiency (EE) [Mbit/Joule]},
        xtick =data,
    legend pos=south east,
    legend style={font=\fontsize{7}{5}\selectfont},
    grid style={line width=.1pt, draw=gray!10},
    major grid style={line width=.2pt,draw=gray!50},
    grid style=dashed,
    grid=both,
  width=\linewidth,
    height=0.8\linewidth
]
    \addplot [mark=x, blue] table [x , y={y2}]{\datatable};
    \addlegendentry{Proposed Method}
    \addplot [mark=o, red] table [x , y={y1}]{\datatable};
        \addlegendentry{MB-SOCP \cite{Van_Chien_Björnson_Larsson_2020}} 
\end{axis}
\end{tikzpicture}
    \caption{The effect of the SE requirement on the energy efficiency of DL transmission of the total system when serving users.}
    \label{fig:eff}
\end{figure}
The energy efficiency metric is the ratio of the network's requested (and achieved) sum rate over the total power consumption. Note that we extrapolate the achieved SE to the system bandwidth $B$ to ensure consistency with the power consumption model of the PA. The power model of the PA considers the entire bandwidth of the system as the PA amplifies the time-domain signal. The energy efficiency metric is then similar to the one proposed in \cite{Ngo_Tran_Duong_Matthaiou_Larsson_2018}: 
\begin{equation}
    \mathrm{EE} = \frac{\sum_{k=1}^{K}B\mathrm{SE}_k}{P_{\mathcal{A}}}.
\end{equation}
Note that the achieved SE at the solution might be higher than the requested SE (cfr. complimentary slackness condition in KKT conditions), yet we only consider the required SE in the energy efficiency metric to be useful. 
Figure~\ref{fig:eff} shows the energy efficiency as a function of the request SE.
Interestingly, the maximum energy efficiency is at 
different points for both methods (1.5 Bps/Hz for our method and 
1.25 Bps/Hz for the baseline). This indicates that our method finds a 
more efficient solution at 1.5 Bps/Hz than at 1.25 Bps/Hz. We find that 
for very low target SE (0.25 Bps/Hz/W), both methods show very low energy efficiency 
due to the users only being allocated sufficient power to 
reach that extremely low threshold. Furthermore, we find a dramatic rise 
in energy efficiency as the target SE increases. This behaviour stems from the slow increase in active APs for the low 
SE targets in Figure~\ref{fig:nbAPs}. 
The fixed power for active serving APs dominates the power consumption at those points. 

\section{Conclusion}
In this paper, we have demonstrated an iterative method for QoS provisioning in Cell-Free Massive MIMO networks.
We have proposed a novel feasibility testing method based on per-user slack variables for a limited subset of active APs. 
This method allows for provisioning the network without turning on every single AP in the network, which is considered to be the case in all state-of-the-art works. 
Our feasibility method provides fine-grained information and allows for using our proposed iterative method for turning on APs. We have shown that our method can exploit the users' spatial information to quickly find solutions that use only slightly more power than extremely costly mixed binary problems from state-of-the-art while eliminating the cost of channel measurement on most APs and is significantly more scalable. In short, we have proposed an algorithm with polynomial complexity scaling for solving an exponentially difficult problem that closely matches the exact solution of that problem via our novel application of slack variables.

\bibliographystyle{ieeetr}
\bibliography{references} 
\end{document}